%% file: text.tex
\documentstyle{europhys} 
\input euromacr.tex

\newcommand{\text}[1]{\hbox{\rm #1}}

\begin{document} 
\euro{}{}{}{} 
\Date{} 
\shorttitle{Y.LEVINSON, DEPHASING IN QUANTUM DOTS} 
\title{Dephasing in a quantum dot due to coupling with a quantum
point contact}
\author{Yehoshua Levinson}   
\institute{Weizmann Institute of Science, 
76100 Rehovot, Israel}

\rec{}{in final form } 

\pacs{  
\Pacs{73}{23.-b}{Mesoscopic systems}
\Pacs{72}{10.-d}{Theory of electronic transport; scattering mechanisms} 
} 

\maketitle 

\begin{abstract}
We consider the dephasing of an one-electron state in a quantum dot
due to charge fluctuations in a biased quantum point contact
coupled to the dot capacitively.
The contribution to the dephasing rate due to the bias depends
on temperature and bias in the same way as shot-noise 
in the point contact at zero
frequency, but do not follow the $|t|^2(1-|t|^2)$ suppression.
\end{abstract} 

Recently, it became possible to measure the phase and hence also the
coherence of resonance electron transmission
through a (pinched) quantum dot (QD) \cite{Y1,Y2,BS,SB}.
Experiments of this type are performed using Aharonov-Bohm rings 
loaded by a QD in one of its arms, the ring being imbedded
in a two- or four-terminal configuration. 
 The conductance of such a system
 contains a term oscillating with magnetic
flux $\Phi$ threaded trough the ring and this term is sensitive
to the phase of the transmission amplitude $t_{QD}$ through the QD.

If the transmission through the QD is coherent
one can calculate ${\cal G}$ from the Landauer formula 
in terms of the transmission through the ring ${\cal T}$.
The transmission amplitude $t_{QD}$ through a pinched QD is small
(it is true even for resonance transmission, since generically
the QD is asymmetric) and for small $t_{QD}$  one finds
$|{\cal T}|^2=|{\cal T}|^2_{0}
+Re \{ a^{*}t_{QD}\exp(2\pi i\Phi/\Phi_{0})\}$,
where $\Phi_{0}$ is the flux quanta.
The two terms correspond to the non-oscillating 
and oscillating parts of ${\cal G}$ in the two-terminal
configuration.
The first is the transmission trough the "free" arm
and the second is due to the interference
between the (multiple) electron paths in the "free" and "loaded" arms.
The effective "amplitude" $a$ is responsible for the geometry
of the ring outside the QD and of the external contacts to the ring.
When the temperature $T=0$ all energy depending 
transmission amplitudes
 has to be calculated at the Fermi energy $\epsilon_{F}$
and for $T\neq 0$ one has to average over thermal energies near
$\epsilon_{F}$. 
 The phase of transmittance trough a QD in an Aharonov-Bohm ring
was analyzed in papers \cite{LB,HW}
and also in connection with the "phase slippage" \cite{OG}.

When the coherence of the electron states in the QD
is destructed due to interaction with some "environment"
the amplitude $t_{QD}$ has to be replaced by its average $<t_{QD}>$
with respect to states of the "environment".
The destruction of coherence is not necessary related to
inelastic scattering \cite{SAY} and this is why we
 call it "dephasing".

One source of such a dephasing can be the capacitive
coupling between the QD and a quantum point contact (PC)
which is close in space to the QD \cite{P,Munp}.
Charge fluctuations in the PC create a fluctuating potential
at the QD and modulate the electron states in the QD.
This modulation being random in time 
dephases the states in the QD and destructs the coherence
of the transmission through the QD.

The charge fluctuations in the PC depend on whether 
the PC is in equilibrium
or supports a current due to an applied bias $V$.
Hence one can expect that the visibility of the ring
conductance oscillations with flux will change with current in the PC.

The aim of this work is to calculate the dephasing time of a state
in a QD induced by its capacitive coupling to a PC.
We will be interested mainly in the additional contribution
to the dephasing rate which is due to the current in the PC.
To simplify the problem we consider dephasing
in an isolated QD not coupled to leads
and coupled capacitively to a PC.

To make things more transparent assume
 only one state in the QD with energy $\epsilon_{0}$
and only one channel in the point contact 
with coordinate $x$ along the channel and an
effective one dimensional barrier potential $U(x)$.
For simplicity we assume the PC
to be symmetric with respect to $x=0$ (the maximum of the barrier).

The Hamiltonian of the QD is 
$H_{QD}=\epsilon_{0}\; c^{+}c$,
where $c$ is an operator removing one electron from the QD state. 

The Hamiltonian of the PC is ($\hbar=1$)
\begin{eqnarray}
H_{PC}=\int dx \psi^{+} (x)
\left[-{1 \over 2m}
{d^2\over dx^2}+U(x)\right]
\psi (x)=\sum_{k} \epsilon_{k}[a^{+}_{k}a_{k}+b^{+}_{k}b_{k}].
\label{PC}
\end{eqnarray}
Here $\psi (x)$ is the electron field operator,
which can be presented as \cite{Bss}
\begin{eqnarray}
\psi (x)=\sum_{k}[a_{k}\chi_{k}(x)+b_{k}\phi_{k}(x)],
\label{ss}
\end{eqnarray}
where
$\chi_{k}(x)$   and
$\phi_{k}(x)$
 are scattering states radiated
from the left and right ohmic contacts   $a$ and $b$
and belonging to the same energy  $\epsilon_{k}$ and
 $a_{k}$, $b_{k}$ are the corresponding
electron operators.
The scattering  states have the following properties ($k>0$)
\begin{eqnarray}
\chi_{k}(x)=L^{-1/2}\{\exp(+ikx)+r_{k}\exp(-ikx)\}, \qquad
{\rm for} \qquad  x\rightarrow -\infty, 
\nonumber\\
\chi_{k}(x)=L^{-1/2}t_{k}\exp(+ikx),   \qquad\qquad\qquad\qquad \;\;
{\rm for} \qquad  x\rightarrow +\infty,
\label{ssa}
\end{eqnarray}
$r_{k}$ and $t_{k}$ being the reflection and transmission coefficients
defined by the potential barrier $U(x)$
 and $L$ is some normalization length.
Similar equations can be written for
 $\phi(x)=\chi(-x)$.

The Hamiltonian of the interaction between the QD and PC,
 assumed to be weak, is
\begin{eqnarray}
H_{int}=
c^{+}c\; W\equiv c^{+}c\;\int dx\; \delta U(x)\psi^{+}(x)\psi(x).
\label{int}
\end{eqnarray}
If one combines $H_{PC}+H_{int}$ one can see that $\delta U(x)$
is the average change of the barrier potential $U(x)$ due to one electron
occupying the QD state when $<c^{+}c>=1$.
This change is localized  near the squeezing of the PC $x=0$.
If one combines $H_{QD}+H_{int}$ one can see that
$W$ is the change of the QD state energy $\epsilon_{0}$
due to the interaction of the electron in the QD
with the electron density in the PC.

The dephasing of the QD state is described by the relaxation
of the average state amplitude $<c(t)>$,
 where $c(t)=\exp(+iHt)c\exp(-iHt)$ is the 
amplitude in Heisenberg representation
with  the total Hamiltonian $H=H_{QD}+H_{PC}+H_{int}$. 

The Heisenberg amplitude equation of motion is
\begin{eqnarray}
(d/dt) c(t)=i\exp(+iHt)[H,c]\exp(-iHt)=
-i[\epsilon_{0}+W(t)] c(t),
\label{cdot}
\end{eqnarray}
with
\begin{eqnarray}
W(t)=\exp(+iHt)W\exp(-iHt).
\label{WH}
\end{eqnarray}

Eq.(\ref{cdot}) demonstrates that
 $W(t)$ is indeed the time depending modulation
of the energy level $\epsilon_{0}$.
Solving this equation and averaging one finds
\begin{eqnarray}
<c(t)>=\left<c(0)\exp(-i\epsilon_{0}t)
 T_{t} \exp \left\{-i\int_{0}^{t}dtW(t) \right\}  \right>,
\label{cav}
\end{eqnarray}
where $T_{t}$ means time ordering.
We decouple the average in Eq.(\ref{cav})
approximating the time ordered 
exponent  by  a Gaussian exponent and obtain
\begin{eqnarray}
<c(t)>=<c(0)>\exp(-i\epsilon_{0}t)\exp(-\Phi(t)),
\label{cg}
\end{eqnarray}
where
\begin{eqnarray}
\Phi(t)={1\over 2}\int_{0}^{t}dt'\int_{0}^{t}dt''K(t'-t''),
\end{eqnarray}
and $K(t)$ is the quantum correlator of the level modulation
\begin{eqnarray}
K(t)={1\over 2}[<W(t)W(0)>+<W(0)W(t)>].
\label{K}
\end{eqnarray}
We also approximate $W(t)$ defined in Eq.(\ref{WH}) by  
\begin{eqnarray}
W(t)=\exp(+iH_{PC}t)W\exp(-iH_{PC}t)
\label{Wt}
\end{eqnarray}
neglecting the influence of the QD on the fluctuations in the PC.
With this approximation the average in Eq.(\ref{K}) reduces to the average
with respect to the state of the PC (in equilibrium or with current)
which plays now the role of an environment.
This last approximation means that we have in mind
 an experimental situation when the current in the QD used
to measure its  linear conductance ${\cal G}$ is small
while the current in the PC can be relatively large.

The correlator $K(t)$ decays in time with some time scale $\tau_{c}$
which is the correlation time of the QD state energy modulation.    
For large $t\gg \tau_{c}$ one find from  Eq.(\ref{cg})
\begin{eqnarray}
<c(t)>=<c(0)>\exp(-i\epsilon_{0}t)\exp(-t/\tau_{\varphi})
\end{eqnarray}
with 
\begin{eqnarray}
{1\over \tau_{\varphi}}={1 \over 2}\int_{-\infty}^{+\infty}dtK(t).
\end{eqnarray}
Hence the {\it average amplitude} of the QD state decays.
Note that under same conditions the population of the QD state is constant
$<c^{+}(t)c(t)>=<c^{+}(0)c(0)>$,
since $H_{int}$ commutes with the number of electrons in the QD $c^{+}c$.
Same is true for the energy of the QD $\epsilon_{0}c^{+}c$.
This means that the decay of $<c(t)>$ with the time constant $\tau_{\varphi}$
is indeed {\it dephasing} and not {\it energy relaxation} or
{\it escape} from the QD.

Using  Eq.(\ref{int}) and  Eq.(\ref{ss}) one finds the
energy modulation in terms of scattering states
\begin{eqnarray}
W=\sum_{k,k'}[a^{+}_{k}a_{k'}A_{k,k'}+b^{+}_{k}b_{k'}B_{k,k'}+
            a^{+}_{k}b_{k'} C_{k,k'}+b^{+}_{k}a_{k'}{\bar C}_{k,k'}]
\label{Ws}
\end{eqnarray}
with integrals describing the interaction between the QD and the PC
\begin{eqnarray}
A_{k,k'}=\int dx \delta U(x)\chi_{k}(x)^{*}\chi_{k'}(x),
\qquad
B_{k,k'}=\int dx \delta U(x)\phi_{k}(x)^{*}\phi_{k'}(x),
\nonumber\\
C_{k,k'}=\int dx \delta U(x)\chi_{k}(x)^{*}\phi_{k'}(x),
\qquad
{\bar C}_{k,k'}=\int dx \delta U(x)\phi_{k}(x)^{*}\chi_{k'}(x)
=C_{k',k}^{*}.
\label{ABC}
\end{eqnarray}
Using Eq.(\ref{Ws}) one can find (similar to the calculation
\cite{L} of the current correlator in a PC)
\begin{eqnarray}
<W(t)W>=
\sum_{k,k'}\exp[i(\epsilon_{k}-\epsilon_{k'})t]
\times \qquad \qquad \qquad
\nonumber\\
\{n_{k}^{a}(1-n_{k'}^{a})|A_{k,k'}|^2+
 n_{k}^{b}(1-n_{k'}^{b})|B_{k,k'}|^2+
 n_{k}^{a}(1-n_{k'}^{b})|C_{k,k'}|^2+
 n_{k}^{b}(1-n_{k'}^{a})|{\bar C}_{k,k'}|^2\},
\label{WW}
\end{eqnarray}
$n_{k}^{a}$ and $n_{k}^{b}$ being the occupation numbers for states
with energy $\epsilon_{k}$ in the left and in the right ohmic
contacts of the PC $a$ and $b$ respectively,
\begin{eqnarray}
 n_{k}^{a,b}=[\exp[-(\epsilon_{k}-\mu_{a,b})/T]+1]^{-1},
\end{eqnarray}
where $\mu_{a,b}$ are the electrochemical potentials in the
ohmic contacts $a,b$ and $V=-(\mu_{a}-\mu_{b})/e$ is the applied 
voltage between $a$ and $b$ (assuming $e>0$).

The spectral density of the QD level modulation can be calculated
using  Eq.(\ref{K}) and Eq.(\ref{WW}) as follows
\begin{eqnarray}
S(\omega)={1\over 2\pi}\int_{-\infty}^{+\infty}dt
\exp(i\omega t)K(t)
={1\over 2}\sum_{k,k'}\delta(\epsilon_{k}-\epsilon_{k'}+\omega)
\times \qquad  \qquad  \qquad  \qquad
\nonumber\\
\{n_{k}^{a}(1-n_{k'}^{a})|A_{k,k'}|^2+
 n_{k}^{b}(1-n_{k'}^{b})|B_{k,k'}|^2+
 n_{k}^{a}(1-n_{k'}^{b})|C_{k,k'}|^2+
 n_{k}^{b}(1-n_{k'}^{a})|{\bar C}_{k,k'}|^2\}+
\nonumber\\
\{\omega \rightarrow -\omega \}.\qquad  \qquad  \qquad
\end{eqnarray}

To simplify the spectral density
 assume that the squeezing in the PC 
(i.e. the barrier $U(x)$) is described by only one
length scale $d$, the energy spacing between
the thresholds for different channels in the PC being
 $\Delta\epsilon\simeq 1/md^2 \ll \epsilon_{F}$.
With this assumption the scale in $k$ for
the coefficients $A,B,C$ and ${\bar C}$ is $d^{-1}$.
Assume also that the bias in the PC and the temperature are small
in the sense that $eV,T\ll \Delta\epsilon$.
One can see that with these assumptions
the relevant frequencies $\omega$ are much smaller
than the inverse of the time of flight through the PC
and that the main contribution come from states close to the Fermi level.
Since $|\epsilon_{k}-\epsilon_{k'}|\simeq v_{F} |k-k'| \simeq \omega $
(where  $v_{F}$ is the Fermi velocity)
one finds that for $\omega \ll v_{F}/d \ll \epsilon_{F}$
the relevant  $|k-k'|\ll d^{-1}$
and one can replace in the coefficients $A,B,C$ and ${\bar C}$
the momenta $k$ and $k'$ by $k_{F}$. After this replacement we just skip
the momenta in the notation and note that $|C|^2=|{\bar C}|^2$.

Changing summation over $k>0$ to integration over $\epsilon_{k}$
and introducing a function \cite{EY}
\begin{eqnarray}
F(\omega)=\int \int d\epsilon d\epsilon'
[\delta(\epsilon- \epsilon'+\omega)+\delta(\epsilon- \epsilon'-\omega)]
n_{\epsilon}(1-n_{\epsilon '})=F(-\omega)
\end{eqnarray}
one finds the explicit expression for the spectral density
\begin{eqnarray}
S(\omega)={L^2\over 8\pi ^2 v_{F}^{2}}
[(|A|^2+|B|^2)F(\omega)+|C|^2 (F(\omega+eV)+F(\omega-eV))],
\end{eqnarray}
Subtracting the equilibrium part of the spectral density
(corresponding to $V=0$)
we find the spectral density of the QD energy modulation
due to the current
\begin{eqnarray}
S_{V}(\omega)=\lambda
 [F(\omega+eV)+F(\omega-eV)-2F(\omega)],
\label{sv}
\end{eqnarray}
where we defined a coupling constant
$\lambda=(L^2/ 8\pi ^2 v_{F}^{2})|C|^2$.

According to  Eq.(\ref{sv}) the spectra of the QD energy modulation
due to nonequilibrium charge fluctuations in the PC
is the same as the shot-noise spectra in the PC
\cite{L,EY}.
The typical modulation frequencies are
determined by the energy window near $\epsilon_{F}$ 
where the electron fluxes from the left and right ohmic contacts
$a$ and $b$ of the PC do not compensate, i.e.
$\omega\simeq eV/\hbar$ for high bias  $eV\gg T$
and $\omega\simeq T/\hbar$ for low bias  $eV\ll T$.
One can check that for low enough temperature and bias
$T,eV \ll \Delta\epsilon$ these  frequencies
are indeed smaller than the inverse time of flight.
These frequencies also define
 the correlation time of the QD level
modulation $\tau_{c}\simeq \omega^{-1}$.

However the amplitude of the energy modulation
determined by the coupling constant $\lambda$
depends on the transmittance through the PC in a way
very different from that of the shot noise,
which is proportional to $|t|^2(1-|t|^2)$.
To see it we can rewrite the coupling constant
in terms of changes in $r$ and $t$ introduced by the
perturbation $\delta U(x)$ to the potential $ U(x)$.
Using the  representation of the Green function 
for the Schr\"{o}dinger equation with potential $U(x)$
in terms of scattering states
\begin{eqnarray}
G(x,x')=(mL/ikt)\chi(x_{>})\phi(x_{<}),
\end{eqnarray}
where $x_{>}$ and $x_{<}$ are the larger and the smaller of $x$ and $x'$,
and also the properties of the scattering states
(for a symmetric barrier)
\begin{eqnarray}
\chi^{*}(x)=r^{*}\chi(x)+t^{*}\phi(x),
\end{eqnarray}
one can find
\begin{eqnarray}
\lambda =(8\pi^2)^{-1}|r^{*}\delta t+t^{*}\delta r_{a}|^2=
         (8\pi^2)^{-1}|r^{*}\delta t+t^{*}\delta r_{b}|^2,
\label{cc}
\end{eqnarray}
where $\delta r_{a}$ and $\delta r_{b}$ are the changes of $r$
for waves coming from ohmic contacts $a$ and $b$ and $t$
is the change in $t$.

Let see how the coupling constant depends on the barrier in the PC.
If the barrier $U(x)$ is infinite 
($|t|^2=0$) the functions $\chi$ and $\phi$
do not overlap and it follows from
Eq.(\ref{ABC}) that $|C|^2=0$
 and $\lambda=0$. Same can be obtained from Eq.(\ref{cc})
since zero transmittance of an infinite barrier
can not be changed by a final perturbation
and hence from $t=0$ it follows that $\delta t=0$.
It means that what matters
for the nonequilibrium dephasing in the QD
is not the applied bias in the PC, but the current.
In case of an infinite barrier applied bias does not change 
the electron density and its fluctuations
in the ohmic contacts $a$ and $b$.

If there is no barrier ($|t|^2=1$),
i.e. the quantum point contact is replaced by a quantum wire,
 $\chi$ and $\phi$ are plane waves and one finds
from Eq.(\ref{cc}) and Eq.(\ref{ABC})
\begin{eqnarray}
\lambda ={1\over 8\pi^2 }|\delta r_{a}|^2={1\over 8\pi^2 }|\delta r_{b}|^2=
{1\over 8\pi^2 v_{F}^2}\left|\int dx \delta U(x)\exp (i2k_{F})\right|^2.
\end{eqnarray}

In terms of the spectral density using Eq.(\ref{sv})
 the nonequilibrium contribution to dephasing rate can be written as
\begin{eqnarray}
\tau_{\varphi}^{-1}=\pi S_{V}(0)=2\pi \lambda [F(eV)-F(0)].
\end{eqnarray}
One can easily estimate that this contribution
 $\tau_{\varphi}^{-1}\simeq \lambda eV/\hbar$
for high bias  $eV\gg T$ and
$\tau_{\varphi}^{-1}\simeq \lambda (eV)^2/T\hbar$
 for low bias  $eV\ll T$.
The dependence of the dephasing time on bias and temperature
follows the shot-noise spectral intensity for zero frequency
\cite{L}. 
Note that $\hbar/\tau_{\varphi}$ is much smaller than the rms average
modulation of the QD level 
$<(\delta\epsilon_{0})^2>^{1/2}\simeq <W^2>^{1/2}$
because of the effect of dynamical narrowing, i.e.
$<(\delta\epsilon_{0})^2>^{1/2}\tau_{c}\ll 1$.
The last inequality is valid because of $\lambda \ll 1$.

The limiting expressions for $\lambda$ demonstrates that in spite
of the fact that the dephasing is originated by the same charge fluctuations
that are responsible for shot noise, the dephasing rate does not
follow generally the intensity of shot noise which is proportional
to $|t|^2(1-|t|^2)$.

A similar calculation of the equilibrium part of modulation spectra
results in
\begin{eqnarray}
S_{0}(\omega)=2(\mu+\lambda)F(\omega),
\end{eqnarray}
where
\begin{eqnarray}
2\mu =(L^2/ 8\pi ^2 v_{F}^{2})[|A|^2+|B|^2]=
(8\pi^2)^{-1}[|r^{*}\delta r_{a}+t^{*}\delta t|^2+
                    |r^{*}\delta r_{b}+t^{*}\delta t|^2].
\end{eqnarray}
A cutoff at the time-of-flight frequency $v_{F}/d$ has to
be introduced to this spectra, but the equilibrium
contribution to the dephasing rate defined by $S_{0}(0)$
is independent on this cutoff and equal to
$\tau_{\varphi}^{-1}=2\pi (\mu+\lambda)T $.
(Note that when the thermal current noise at zero frequency
is calculated in this way it is in agreement
 with the d.c. conductance according to the
Nyquist theorem).

If the QD is not pinched and the electron in state $\epsilon_{0}$
has a finite escape rate to the leads $\Gamma$ 
the dephasing rate competes with this rate.
For zero dephasing the (nonaveraged) transmission amplitude
$t_{QD}$ contains a resonant factor
$-i/[(\epsilon_{F}-\epsilon_{0})+i\Gamma]$.
Due to dephasing this factor is replaced in the averaged
amplitude $<t_{QD}>$ by
\begin{eqnarray}
\int_{0}^{\infty}dt\exp [-\Gamma t-\Phi(t)+i(\epsilon_{F}-\epsilon_{0})t].
\label{df}
\end{eqnarray}
Generally the factor Eq.(\ref{df}) defines a non Lorentzian line-shape
but in the case of dynamical narrowing 
$<(\delta\epsilon_{0})^2>^{1/2}\tau_{c}\ll 1$
it reduces to
$-i/[(\epsilon_{F}-\epsilon_{0})+i(\Gamma+\hbar/\tau_{\varphi})]$
just adding the dephasing rate to the escape broadening.

Recently the problem of dephasing was also addressed
in an unpublished paper of Aleiner, Wingreen and Meir \cite{AWM}
using a different approach.

\stars

This work was initiated as a result of discussions with
M.Heiblum and E.Buks 
about the possibility of the so called "Which Path?" experiment
in the Aharonov-Bohm interferometer. I acknowledge these
discussions very much.
I am thankful to Y.Imry for discussions related to 
 level broadening and dephasing rate
 and also to A.Stern and F.von Oppen
 for discussing the results of this work.

\vspace*{-4mm} 
\newcommand{\noopsort}[1]{} \newcommand{\printfirst}[2]{#1} 
  \newcommand{\singleletter}[1]{#1} \newcommand{\switchargs}[2]{#2#1}

\end{document}

%% file: euromacr.tex
\def\And{{\rm and\ }}

\def\stars{\bigskip\centerline{***}\medskip}

\newif\ifboo \boofalse

\def\Review#1{\boofalse{\it #1},}
\def\Name#1{{\sc #1},}
\def\Vol#1{\ifboo Vol. {\bf #1}\else{\bf #1}\fi}
\def\Year#1{\ifboo #1\else(#1)\fi}

\def\Page#1{\ifboo {\rm p. #1}\else{\rm #1}\fi}